\documentclass[twocolumn,showpacs,preprintnumbers,amsmath,amssymb,a4paper]{revtex4}
\usepackage{graphicx}% Include figure files
\usepackage{epsfig}
\usepackage{dcolumn}% Align table columns on decimal point
\usepackage{bm}% bold math
\usepackage{subfigure}
\graphicspath{{converted_graphics/}}
\begin{document}

%\preprint{APS/123-QED}

\title{Ion-beam modification of the magnetic properties of GaMnAs epilayers}%

\author{E. H. C. P. Sinnecker}
\author{G. M. Penello}
\author{T. G. Rappoport}
\author{M. M. Sant'Anna}
\author{D. E. R. Souza}
\author{M. P. Pires}

\affiliation{Instituto de F\'{\i}sica, Universidade Federal do Rio de Janeiro, Rio de Janeiro, RJ, 21941-909, Brazil}
\author{J. K. Furdyna}
\author{X. Liu}
\affiliation{Department of Physics, University of Notre Dame, Notre Dame, IN 46556, USA}%


\date{\today}% It is always \today, today,

\begin{abstract}
We study the controlled introduction of defects in GaMnAs by irradiating the samples with energetic ion beams, which modify the magnetic properties of the DMS. Our study focuses on the low-carrier-density regime, starting with as-grown GaMnAs films and decreasing even further the number of carriers, through a sequence of irradiation doses. We did a systematic study of magnetization as a function of temperature and of the irradiation ion dose. We also performed in-situ room temperature resistivity measurements as a function of the ion dose. We observe that both magnetic and transport properties of the samples can be experimentally manipulated by controlling the ion-beam parameters. For highly irradiated samples, the magnetic measurements indicate the formation of magnetic clusters together with a transition to an insulating state. The experimental data are compared with mean-field calculations for magnetization. The independent control of disorder and carrier density in the calculations allows further insight on the individual role of this two factors in the ion-beam-induced modification of GaMnAs.
\end{abstract}

\maketitle

\section{Introduction}

During the last decade, III-V diluted magnetic semiconductors (DMS) such as GaMnAs have been the subject of intensive research, specially due to their possible applications in semiconductor-based spintronics. Ga$_{1-x}$Mn$_x$As presents long range magnetic order between spaced Mn ions, mediated by holes~\cite{Vanesch1997,Jungwirth2006,Dietl2008}. The existence of these holes in the band structure of the material is undoubtedly due to the incorporation of Mn ions. 
Consequently, crystalline defects have a crucial role in both electrical and magnetic properties of ferromagnetic semiconductors and in device applications~\cite{Macdonald2005}. Continuous refinement of growth techniques has led to a significant improvement of the upper limit for the  incorporation of Mn atoms at appropriate sites. This improvement towards the high-carrier-density regime has been very effective to enhance the Curie temperature of Ga$_{1-x}$Mn$_x$As~\cite{Chiba2007}. On the other hand, some experiments studying and proposing prototypes of actual spintronic devices also point to the importance of the low-carrier-density regime~\cite{Chiba2008}. This is the case, for instance, of the proposed control of GaMnAs magnetic anisotropy by the application of electric fields~\cite{Chiba2008}. 

Different paths have been used to experimentally access the effects of carrier density and structural disorder while keeping the Mn concentration constant. Annealing-induced changes are very effective to enhance Curie temperature and are used to control magnetization and magnetic anisotropy~\cite{Potashnik2001,Stanciu2005,Chiba2007}. However, it is not trivial to quantify the effect of a given annealing procedure on the density of defects. More recently, the addition of hydrogen to the epilayer has been used to change GaMnAs magnetization properties~\cite{Goennenwein2004,Thevenard2007}. In this latter case, it is possible to quantify the creation of defects by estimating the amount of hydrogen aggregated to GaMnAs. A third path is the implantation of heavy ions in the GaMnAs sample. Beams of Ga$^+$ of 30 keV have been used to introduce deep trap levels in the epilayer~\cite{Kato2005}. In the two latter methods additional material is incorporated in the GaMnAs sample.  We use a different approach to study structural defects in GaMnAs films.

We produce defects in the GaMnAs sample in a controlled way by irradiating the samples with ion beams fast enough to cross the GaMnAs layer. Thus, the projectile ions are implanted only in the non-magnetic GaAs substract. However, going through the GaMnAs film, the fast ion beam leaves behind defects in the crystalline structure, enhancing the disorder of the system. The same technique has been applied in the study of radiation tolerance of semiconductors and devices that are used in solar cells, detectors and satellites. Consequently, there is a great amount of information on the characterization of defects produced by energetic ion beams.  They have also been used in different contexts: the study of the effects of structural disorder in superconductors~\cite{Weaver1992,Paulius1997,Bugoslavsky2001,Analytis2006}, ferromagnets~\cite{Kihkonent1992}, n-type and p-type~\cite{Stievenard1986} semiconductors.

In this article, we study the manipulation of carrier density and structural disorder of GaMnAs by the use of MeV ion beams. We focus on the low-carrier-density regime, decreasing this density as defects are introduced by the ion beam. A second effect of the ion-beam irradiation is the increase of structural disorder controlled by the irradiation dose. We present magnetic and transport measurements of GaMnAs epilayers irradiated with energetic ion beams and analyze the effect of the irradiation on the magnetic and electric properties of the DMS.

The article is organized as follows: in section \ref{sec1} we discuss the irradiation technique, the defects normally created in semiconductors by this process and the method that can be used to quantify the formation of defects in the DMS due to the irradiation. X-ray diffraction measurements are used to discuss the enhancement of the density of defects in both the epilayer and in the substrate. In section \ref{sec2} we present room temperature sheet resistance measurements as a function of the ion dose and discuss the similarities between GaMnAs and other p-doped GaAs semiconductors. In section \ref{sec3}, we present the magnetic measurements performed in samples with different irradiation doses and discuss how the irradiation process affects the magnetic order in GaMnAs. Finally, in section \ref{sec4} we use a mean-field approach for an impurity band model to compare our experimental results with the theoretical magnetization curves and understand the roles of the decrease of carrier concentration and of the enhancement of the structural disorder in the modification of the magnetic properties of the DMS.

\section{Irradiation process\label{sec1}}

In n-type and p-type  GaAs semiconductors, the beams introduce a rich variety of  defects. The defects in n-type GaAs are better understood than the ones created in p-type GaAs.  However, it is well established that in both cases  these defects in GaAs semiconductors reside in the As sublattice and most of them are primary defects (related to vacancies and interstitials)~\cite{Pons1985}. The irradiation process produces similar quantities of defects in the Ga and As sublattices but the vacancies in the Ga sublattice tend to recombine immediately with the interstitials, since they have opposite charge. We do not have any previous information on the role of Manganese atoms in the formation of complex defects in irradiated GaMnAs. However, previous studies of such defects in non-irradiated samples suggest the possibility of formation of isolated Mn interstitials with As nearest neighbors and pairs of Mn interstitials with As nearest neighbors~\cite{Stroppa2007,Sullivan2003}.

The range of the implanted-ion profile is determined by the choices of projectile atomic number (Z) and kinetic energy. This range is independent of the beam dose (measured in ions/cm$^2$). On the other hand, in order to determine the density of the ion-induced defects in the relevant sample epilayer, knowledge of the dose is also necessary. Ion current density and irradiation time are the parameters used to experimentally control the beam dose. 

The Ga$_{1-x}$Mn$_x$As samples are grown on a GaAs substrate by Low-Temperature Molecular Beam Epitaxy (LT-MBE). The epilayers have 200nm or 100 nm thickness and a Mn concentration, $x$, of 5\%. Hall measurements performed with non-irradiated samples determined a density of holes of 7$\times 10^{19}$ cm$^{-3}$ for the 200nm and 2$\times 10^{19}$ cm$^{-3}$ for the 100 nm samples. We irradiate the samples, at room temperature, with ion beams from a 1.7 MeV NEC Tandem Accelerator.  As the Z of the used ion increases, the dose necessary to create the same density  of defects decreases. For decreasing energies, there is an increase in the creation of defects. In order to cover a large range of defect densities, we use 100 keV protons (4 different doses), 1000 keV protons (3 different doses) and  700 keV Li$^+$ ions (4 different doses). Ion-beam current densities range from 0.21 pA/cm$^2$ to 1 nA/cm$^2$ and ion-beam doses are in the range between 1.3$\times 10^9$ ions/cm$^2$ and 6.7$\times 10^{14}$ ions/cm$^2$.

Both the ion beam implantation profile and the vacancy density profile are simulated using the SRIM 2008 code~\cite{srim}. The complexity of the cascade collisions in the solid target limits the accuracy of the simulation for defect production. However, reasonable agreement between SRIM simulation and experiment has been obtained for the vacancy production in GaAs~\cite{khanna2005}. We used 15 eV for the displacement parameter in the SRIM code. Although there are no previous simulations for GaMnAs, this value is well accepted for GaAs simulations~\cite{khanna2005}. 

\begin{figure}[h]
\includegraphics[width=1.00\columnwidth]{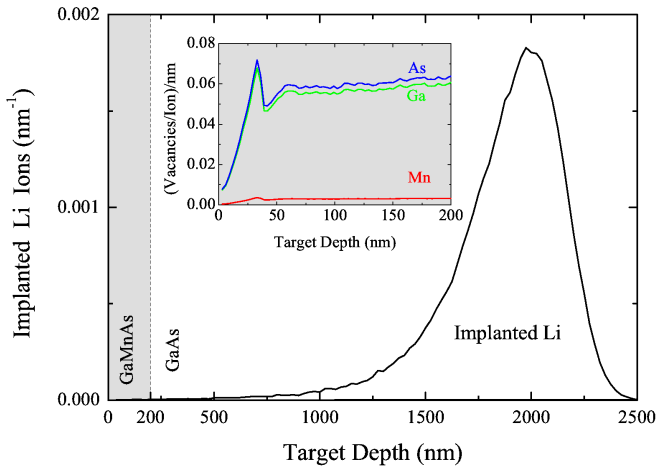}
\caption{The depth profile of 700 keV Li$^+$ ions implanted in the target. A negligible fraction of incident ions is implanted in the magnetic GaMnAs epilayer. The inset shows the number of created vacancies / Li$^+$ ion / nm, for each target element (Ga, Mn, and As), as a function of the epilayer depth. Simulations obtained with the SRIM 2008 code.\label{Fig1}}
\end{figure}

Figure~\ref{Fig1} shows the SRIM simulation results for 700 keV Lithium projectiles. It shows that almost none of the projectiles are implanted into the GaMnAs. The inset at Fig.~\ref{Fig1} indicates a relatively uniform defect creation through the epilayer. The integral of each curve of the inset in Fig.~\ref{Fig1} gives the number of vacancies per ion created in the epilayer for the correspondent target component. These numbers, multiplied by the beam dose and divided by the epilayer thickness, give the average volume densities of produced crystal vacancies for Ga, Mn, and As. We use these densities to quantify the induced disorder in the GaMnAs epilayer. Although SRIM has to rely on theoretical simulations of the motion of atoms in the irradiated sample, this procedure still gives a quantitative scale for the introduced disorder that temperature changes can not provide. 
 
It is important to point out that SRIM simulations do not consider temperature and do not account for recombination processes. This can be clearly seen in Fig.~\ref{Fig1}, where the simulation shows a considerable number of Ga vacancies created by the irradiation, although we know that most of these defects recombine. More specific molecular dynamics simulations for irradiated GaMnAs, like those reported for GaAlAs~\cite{AlGaAs}, would be desirable but are not available in the literature. However, the effect of this recombination is roughly dose independent, except for extremely large doses that can cause amorphization even in the GaMnAs layer. Thus, recombination processes would basically change our chosen defect scale by a constant factor.

\subsection{X-ray diffraction of ion-beam irradiated samples\label{subsecxrays}} 

X-Ray Diffraction (XRD) is routinely used to characterize MBE-grown Ga$_{1-x}$Mn$_x$As epilayers (e.g.~\cite{Zhao2005, Daeubler2006, Bihler2008} ). These measurements provide estimates for changes in lattice parameters relative to bulk GaAs and, therefore, for Mn content.  Furthermore, XRD is also a tool that has been used to experimentally access the effects of ion-beam irradiation on bulk GaAs samples ~\cite{Wie1986, Xiong1991}. In this latter case, XRD measurements and data analysis based on dynamical diffraction theory result in information on the related depth profiles for implanted ions, induced structural defects, and strain. Our irradiated samples combine characteristics of both kinds of samples mentioned above: non-irradiated GaMnAs samples and bulk implanted GaAs. Thus, it is convenient to use XRD measurements to pave the way for the transport and magnetization studies of irradiated GaMnAs epilayers presented in the following sections of this paper. 

\begin{figure}[h]
\includegraphics[width=1.00\columnwidth]{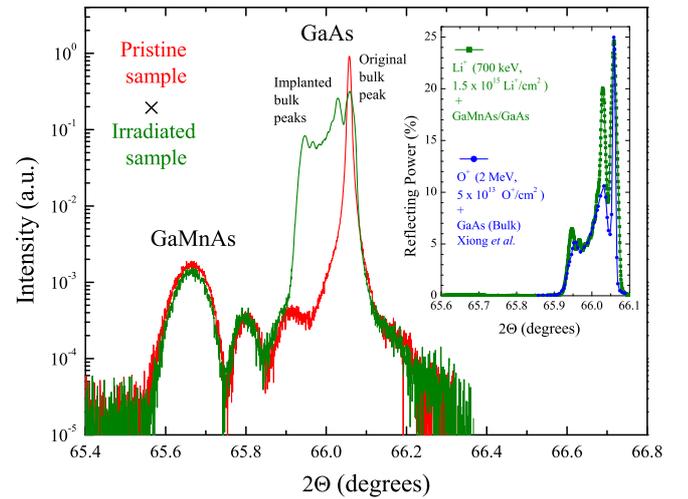}
\caption{Comparison of measured $2\Theta/\Omega$ scans for symmetric (004) reflex of 100 nm Ga$_{1-x}$Mn$_x$As epilayers, with $x$=0.05, grown on GaAs. The angle 
2$\Theta=66.048^\circ$  corresponds to the (004) reflex of the GaAs substrate. The curve with a narrow GaAs bulk peak corresponds to pristine sample. The curve with a set of peaks for GaAs correspond to a sample irradiated with 700 keV Li$^+$ ion beam, with a dose of 1.5$\times 10^{15}$ Li ions/cm$^2$. The inset shows a comparison between our data for 700 keV Li$^+$ and data for a projectile with equivalent implantation range, 2 MeV O$^+$ ions, incidend on GaAs bulk sample~\cite{Xiong1991}. \label{Fig2}}
\end{figure}

Figure~\ref{Fig2} shows X Ray Diffraction Rocking Curves for pristine and for irradiated 100 nm GaMnAs epilayers grown on top of GaAs. The data was taken by a Bede 200 x-ray diffractometer equipped with an Enraf Nonius x-ray generator. A copper tube source was selected, and the x-ray Ka line was filtered using a Ni foil.Like in all our samples, there is also a 100 nm LT-GaAs layer between the GaMnAs layer and the GaAs Bulk. The irradiation dose is 1.5$\times 10^{15}$ Li ions/cm$^2$ and the Lithium beam energy is 700 keV. 
The pristine sample shows in Fig.~\ref{Fig2} the typical behavior of MBE-grown Ga$_{1-x}$Mn$_x$As epilayers. There are two main structures. The larger and narrower peak corresponds to the GaAs substrate (with a merged contribution from the 10 nm  LT-GaAs layer~\cite{Zhao2005}) and a smaller and wider peak corresponding to the GaMnAs epilayer. The other structures are interference fringes. Those fringes can actually be used to estimate the width of the GaMnAs epilayer (100 nm in this case). 

Figure~\ref{Fig2} also shows an additional set of peaks characteristic of Rocking Curves for ion-beam irradiated GaAs bulk samples. Deep into the sample, far from the GaMnAs epilayer, Li ions are implanted (see Fig.~\ref{Fig1}) leading to an approximately gaussian depth profile of cristaline defects. This results in a continuous modification of the GaAs lattice parameter ~\cite{Wie1986, Xiong1991} with a maximum strain modification around 2000 nm. The boundaries of this buried layer of defects are not as sharp as in the case of the MBE-grown Ga$_{1-x}$Mn$_x$As. However, the implanted layer of defects is well enough defined to result in a second set of fringes of interference (see Fig.~\ref{Fig2}). The effective width of this layer can also be estimated from the distance between the fringes and is approximately 650 nm. This value is consistent with the implantation depth profile shown in Fig.~\ref{Fig1}and obtained from SRIM simulation. In the inset of Fig. 2 we compare, in a linear scale, Rocking Curves for our irradiated sample and for bulk GaAs irradiated by 2 MeV oxygen ions ~\cite{Xiong1991}. The 700 keV Li$^+$ ions and the 2 MeV O$^+$ ions have approximately the same implantation range and result in similar bulk Rocking Curves. It is important to note, however, that the amount of defects (and therefore strain) introduced near the surface is much smaller than in the bulk. The previous study of Xiong et al with O$^+$ beam on bulk GaAs~\cite{Xiong1991} also shows that, even for depths around the maximum of deposition the Rocking curves shown at the inset of Fig.~\ref{Fig2} correspond to only a moderate dose in terms of the density of defects created. An implantation dose 200 times larger would be necessary to create an amorphized layer of GaAs in the bulk~\cite{Xiong1991}.  

Regarding the GaMnAs peak, no change in position or witdth, within experimental errors, is noticed from pristine to irradiated sample in Fig.~\ref{Fig2}. The unaltered diffraction angle shows negligible changes in lattice parameters of the GaMnAs layer. Previous studies of ion-beam irradiated 2 mm GaN epilayers ~\cite{Puviarasu2009} show a small increase of the peak width attributed to the enhancement of defect density. In our irradiated sample such a modification is not observed.
In summary, the XRD measurements show that the ion-beam irradiation does not change appreciably the lattice constants of our GaMnAs epilayers. The passage of the ion beam  through the superficial layer moves part of the atoms from their initial positions. A large variety of defects is created. The average rearrangement, however, does not result in a relevant change of the Ga$_{1-x}$Mn$_x$As crystalline structure.  On the other hand, this rearrangement does modify transport and magnetic properties of the epilayer. Particular defects, like the transformation of a substitutional Mn in an interstitial Mn, are certainly more decisive than others in the modification of these properties. In the next two sections we discuss measurements that can access those ion-beam induced changes, and the correspondent experimental-technique limitations in this specific case.

\section{In-situ sheet resistance measurements\label{sec2}} 

To facilitate the comparison between the irradiation process on GaMnAs and  
results on p-GaAs, we performed  sheet resistance $R_s$ measurements. The $R_s$ 
measurements are routinely used to characterize doped non-magnetic semiconductors 
modified by light energetic ion beams in the so-called isolation
 process~\cite{Souza1996,Carmody2003}. In that case, the use of ion beams induce a dramatic decrease of carrier concentration in selected area sectors of the
 epilayer. This procedure keeps the planarity of the sample
 and is used as a tool in the microelectronics industry.  

In order to calibrate our equipment, we performed a first irradiation of a GaAs sample $\delta$-doped with C. We compared the doses necessary to electrically isolate the sample with the ones of Ref. [\onlinecite{Boudinov}] and their results were reproduced. Subsequently we performed the irradiation of the DMS samples with different ions and energies. Combining the different projectiles we span 6 orders of magnitude in the introduced defect density. The irradiation by 1000 keV protons with doses up to 6.7$\times 10^{14}$ ions/cm$^2$ did not produce any considerable change in the properties of the DMS. High doses of 100 keV proton produced a very small change. The choice of Li$^+$, an ion with larger atomic number and mass, finally produced visible changes in the magnetic and transport properties of the samples. 

\begin{figure}[tbp] % float placement: (h)ere, page (t)op, page (b)ottom, other (p)age
  \centering
    \includegraphics[width=1.00\columnwidth]{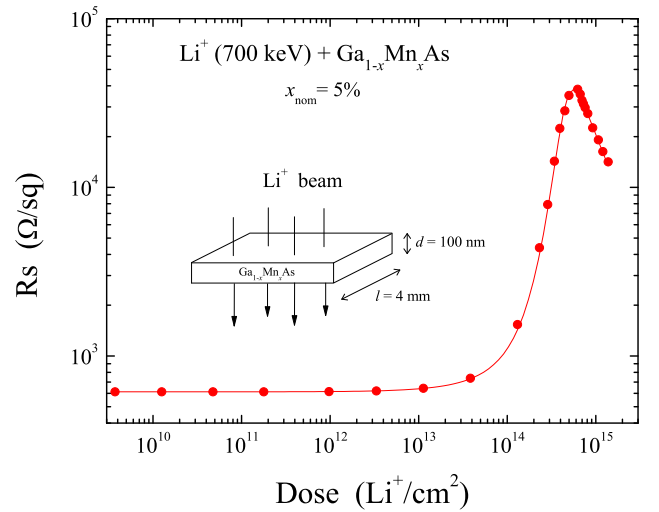}
  \caption{Sheet resistance (SR) measured as a function of the 700 keV Li$^+$ ion dose for a 100 nm thick GaMnAs sample.}
  \label{Fig3}
\end{figure}
In this experiment, the $R_s$ values were measured in-situ after every irradiation step
 of the dose accumulation. The measurements were performed at room temperature using the van der Pauw method~\cite{VandePawl}. Figure \ref{Fig3} shows the sheet resistance as a
 function of the 700 keV Li$^+$ ion dose for a 100 nm thick GaMnAs sample. Different behaviors can be 
identified according to the range of doses analyzed. 

At very low doses $R_s$ increases slowly. As the dose increases, there is a pronounced enhancement of the $R_s$ up to a maximum and then a decrease with a power law of the dose. This behavior is also found in non
-magnetic epilayers of both n and p doped semiconductors grown on top 
of undoped wafers~\cite{Carmody2003}. Bellow the maximum of Fig. \ref{Fig3}, the 
increase of the $R_s$ is a direct consequence of the decrease of the density of carriers
 in the epilayer caused by crystalline defects introduced by the trespassing ion beam.
 To understand the high-dose regime, however, it is necessary to take into account what
happens to the bulk. The Li ions are buried in the GaAs substrate at a depth of approximately 2000 nm (see Fig.~\ref{Fig1}). For  high ion doses they create a deep layer of modified GaAs in the bulk  with a concentration of defects so high that the conduction by hopping between GaAs defects becomes relevant. Therefore, a simple model (e.g.
 Ref.~\cite{Boudinov}) considering two resistances in parallel (the GaMnAs sheet at surface and the high-defect-density GaAs sheet deep at 2000 nm) can explain the dose dependence
 seen in Fig.~\ref{Fig3}.

It is important to note one peculiarity of the effect of incident ion beams in the case 
of the GaMnAs epilayer: the buried implanted layer has no magnetic properties. Thus,
 magnetic measurements, which infer the hole-mediated ferromagnetism are expected to
 reflect a smooth decrease in carrier densities coming solely from the GaMnAs epilayer.
On the other hand, additional effects, as the increase of structural disorder, may be induced by the ion beams and compete with carrier concentration changes to modify the
 magnetic properties of the DMS.

\section{Magnetization measurements\label{sec3}}

For the magnetization measurements discussed in this section, we used a 200 nm thick sample with a higher concentration of carriers than the one in section \ref{sec2} but the same nominal concentration of Mn.  After each ion irradiation, we perform magnetic measurements in a SQUID magnetometer. In order to analyze the sample anisotropy, an in-plane magnetic field is applied on the sample with two orthogonal orientations: parallel and perpendicular to easy axis ([110]) of the non-irradiated sample.

\begin{figure}[b]
\includegraphics[width=0.9\columnwidth]{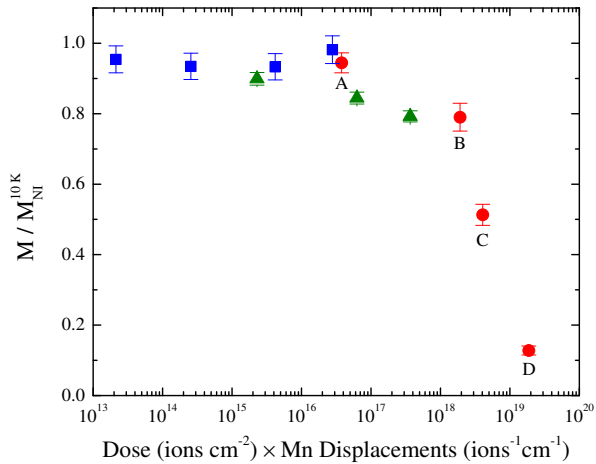}
\caption{Magnetization at 10 K versus produced Mn displacement density for different
samples and ion beams: 1 MeV H$^+$, squares; 100 keV H$^+$,
triangles; 700 keV Li$^+$, circles. The figure shows an universal
behavior where the magnetization is independent of the atomic number
and energy of the ion. The labels A-D refer to the 200 nm thick sample with increasing irradiation doses of Li$^+$. \label{Fig4}}
\end{figure}

In Fig.~\ref{Fig4} we show the value of the [110] magnetization at 10 K as a function of the number of created Mn vacancies obtained by the SRIM simulation, which also measures the amount of disorder in the system. Here, we clearly see a universal curve, which is independent of the type or energy of the ion beam. The total density of Mn vacancies is used here as a scale. The SRIM simulations do not take into account recombination processes and these numbers can be considerably smaller than the ones shown here.

\begin{figure}[h]
\includegraphics[width=1.0\columnwidth]{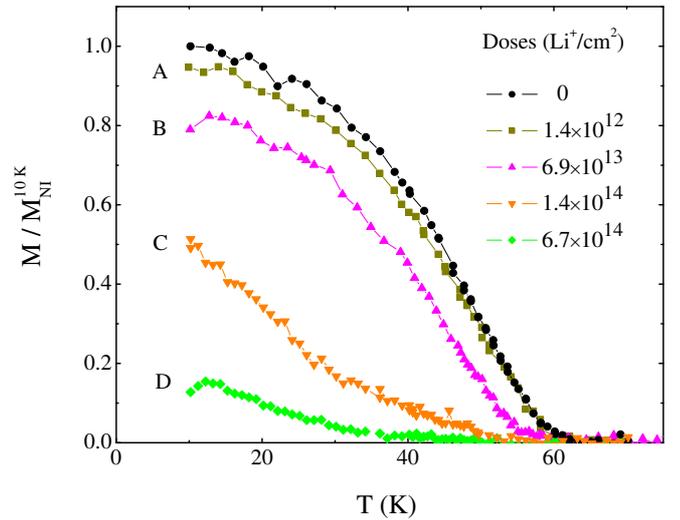}
\caption{ Magnetization versus temperature for samples irradiated
with Li$^+$. The magnetization is normalized by the value of the
non-irradiated sample at 10 K. The figure shows that the
magnetization is gradually suppressed with the increase of the ion
dose. \label{Fig5}}
\end{figure}

Figure \ref{Fig5} shows the easy-axis magnetization versus temperature measurements for the samples irradiated with Li$^+$. We can clearly see a decrease in the magnetization at low temperatures, even though the critical temperature is only slightly modified.  Up to now, most of the discussion was focused on the consequences of the variation of resistivity due to the irradiation process but we can also understand the role of the increase of disorder in GaMnAs. For an increasing amount of disorder, we observe a change in the concavity of the magnetization curves (see Fig.\ref{Fig5} ), which is compatible with highly compensated GaMnAs samples. The suppression of the magnetization for low T is not followed by considerable variation of $T_c$.  Previous theoretical calculations predict that for a fixed number of carriers, an increase of disorder will lead to a decrease of the magnetization at low temperatures and a considerable {\it increase of $T_c$}~\cite{Berciu2001,Timm2002}. Here, any variation of the amount of disorder is accompanied by a change of the resistivity, as can be seen in Table~\ref{tab1} and consequently, the carrier concentration. As a decrease of the carrier concentration decreases $T_c$, these two effects compensate each other and $T_c$ remains almost unaltered until very high doses of ion beams.

\begin{table}
\caption{Properties of GaMnAs sample irradiated with 700 keV Li$^{+}$ ions: created Mn Vacancies (SRIM simulation) and sheet resistance (at room temperature)}

\begin{tabular}{c c c c}
\hline\hline
Irradiation & Dose & Mn$_{vac}$ & sheet resistance\\
 & (Li$^+$/cm$^2$)& (cm$^{-3}$) &(ohm/sq)\\
\hline
non-irrad & 0 & 0  & 432\\
C & $1.4\times10^{14}$ & $4.0\times10^{18}$  & 822 \\
D & $6.7\times10^{14}$ & $1.9\times10^{19}$  & 3266\\
\hline
\hline
\end{tabular}
\label{tab1}
\end{table}

\begin{figure}[b]
\includegraphics[width=1.00\columnwidth,clip]{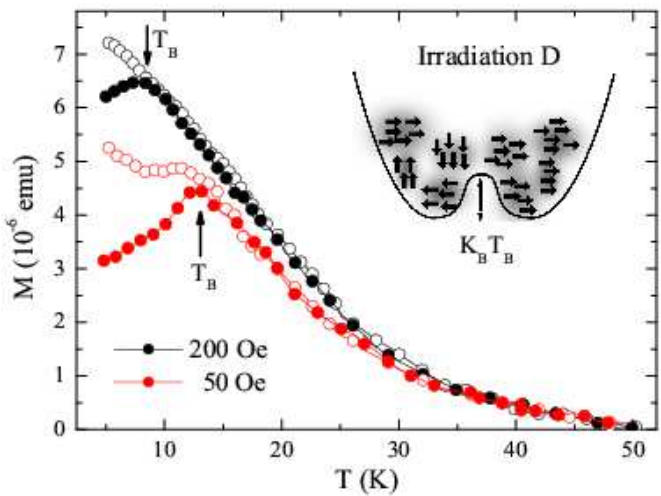}
\caption{(a) ZFC (closed symbols) and FC (open symbols) magnetization $\times$ temperature measurements for two different applied fields: 50 Oe and 200 Oe. Inset: Schematic representation of the blocking temperature barrier separating a configuration of random oriented magnetic clusters from a configuration of magnetically oriented clusters\label{Fig6}}
\end{figure}

In  Fig.~\ref{Fig5}, for the most intense irradiation (D) we see an unusual decrease of the magnetization for low temperatures. In principle, this feature could be attributed to the creation of shallow levels in the DMS where the trapped carriers can be thermally activated~\cite{Calderon2007}. Another possibility is the existence of a blocking temperature due to the formation of magnetic clusters in highly irradiated samples. To further investigate these possibilities, we performed standard zero-field cooled (ZFC) and field-cooled (FC) magnetic measurements. Fig.~\ref{Fig6} shows that the ZFC and FC measurements have different behaviors at low temperature. These results rule out the picture of an increase in the density of holes due to thermally activated carriers. The existence of a blocking temperature $T_B$ that depends on the applied field is consistent with the existence of magnetic clusters.  After the irradiation process, due to the reduced number of carriers and disorder, the holes are localized, similarly with what is observed for GaMnAs films at low Mn concentration~\cite{science2010}.  Consequently, the DMS is composed of weakly interacting clusters, probably close to the percolation transition. For low temperatures, the ZFC measurements can freeze the system in a random orientation of these ordered magnetic clusters. The system needs to overcome a barrier  of $k_BT_B \sim 0.1$ meV in order to reorientate the magnetic clusters, as illustrated in the inset of Fig.~\ref{Fig6}.  Similar behaviour has been predicted by Monte-Carlo simulations~\cite{Mayr2002}.
\vskip 0.2 cm

\begin{figure}[tbp] % float placement: (h)ere, page (t)op, page (b)ottom, other (p)age
  \centering
   \includegraphics[width=1.0\columnwidth]{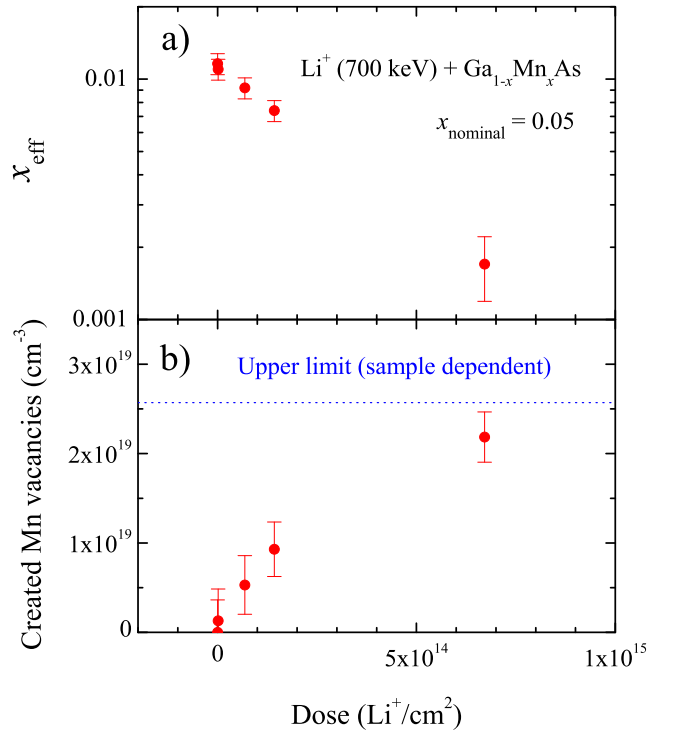}
  \caption{Parameters experimentally determined from magnetization measurements, as a function of the incident 700 keV Li$^+$ ion dose: (a)  the effective value of $x$ in Ga$_{1-x}$Mn$_x$As (b) the density of Mn vacancies created by the ion beam. Due to the formation of magnetic clusters in the highly-irradiated sample (see Fig.~\protect\ref{Fig6}), the estimate of the effective concentration for this particular experimental point is less accurate and the quantitative estimate is less reliable.} 
  \label{Fig7}
\end{figure}

We can further analyze the effect of the irradiation process by estimating the fraction of Mn atoms that are participating in the magnetism. The effective value $x_{eff}$ in Ga$_{1-x}$Mn$_x$As is the concentration of Mn atoms contributing to the magnetic order. If we make the assumption that the interstitial Mn are paramagnetic and do not contribute to the measured magnetic moment, we can estimate   $x_{eff}$ from the saturation magnetization $M_s$ using $M_s=4x_{eff}g\mu_BS/V_{\mbox{cell}}$ where $g$ is the gyromagnetic ratio, $\mu_B$ is the Bohr magneton, $V_{\mbox{cell}}$ is the volume of the unit cell and $S=5/2$ is the spin of Mn.  It is important to notice that here we consider the concentration of of positional Mn as the $x_{eff}$. For high nominal concentrations of Mn, this assumption is less accurate, as part of the interstitial Mn are coupled antiferromagnetically and $x_{eff}$ is smaller than the concentration of positional Mn~\cite{Wang2004}.

Figure \ref{Fig7}(a) shows values of $x_{eff}$ obtained from magnetization measurements as a function of the dose of incident 700 Kev Li$^+$ ions. The irradiation, on average, progressively removes Mn atoms from sites where they are active for ferromagnetism. Although the nominal value of $x$ for this sample is 5 \%, the $x_{eff}$ before irradiation is only 1.2 \%, a difference that can expected for as-grown samples~\cite{sadowski2009}. For the highest dose irradiation, $x_{eff}$ decreases by about one order of magnitude. However, the estimation of the effective concentration in this sample is less accurate once Fig.~\ref{Fig6} shows different values of the magnetization at low temperatures for ZFC and FC measurements. It is possible that for the saturation magnetization we still have isolated magnetic clusters with different magnetic moment orientations. Although more Mn participate in the magnetism, this is not reflected in the total magnetization of the system. Magnetic frustration is also a source of the underestimation of $x_{eff}$~\cite{zarand2002}.

In order to compare these experimental results with computational simulations like those using the SRIM code~\cite{srim}, it is convenient to present the data in an alternative way. Figure \ref{Fig7} (b) presents the density of active Mn atoms removed from their initial sites  as a function of the dose of incident ions. The derivative of this curve provides the rate of introduction of defects, Ti, (i.e., number of created Mn vacancies per incident ion per unit of length). This quantity is also an output from SRIM calculations and is sometimes used as an input parameter for theoretical modeling of the effects of ion beams in electrical properties of solids. This modeling often assumes that the variation of concentration of a certain kind of defect depends linearly on the implantation dose.

Our defect-specific experimental data (for Mn vacancy creation) offer a way to benchmark calculations of Ti and to test the above-mentioned linearity assumption. In Fig. \ref{Fig7}(b) the linear regime can be seen for doses up to approximately 1.5 $\times 10^{14}$ Li$^+$ ions/cm$^2$. This behavior obviously has to change for higher doses because otherwise the number of displaced Mn atoms would increase beyond the total number of Mn atoms. Our measurement at 6.7 $\times 10^{14}$ Li$^+$ ions/cm$^2$ is clearly beyond the linear regime. Close to the origin, corresponding to x$_{eff}$=1.2 \%, we experimentally obtain Ti= 6.1 $\times 10^{-3}$ Mn vac/(ion angstrom). The SRIM calculation, assuming displacement energy of 15 eV for all atoms, give us Ti= 8 $\times 10^{-3}$ Mn vac/(ion angstrom). The SRIM code simulations and our estimates therefore agree within 30\%.

\section{Hole concentration $\times$ structural disorder\label{sec4}} 
An effective approach to understand the effect of the irradiation process is to compare our experimental results with theoretical calculations that allow for the variation of carriers and spin concentration and also take into account the structural disorder. Since we are dealing with small carrier and Mn concentrations in as-grown samples,  an impurity band model is better suited for this analysis~\cite{Alvarez2003, Burch2006, Jungwirth2007}. The Hamiltonian is given by

\begin{equation}
\label{eq1}
\nonumber {\cal H}=\sum_{i,j}^{}t_{ij}c^{\dagger}_{i\sigma}
c_{j\sigma} +\sum_{i,j}^{} J_{ij} \vec{S}_i\cdot \vec{s}_i
-g \mu_BH\sum_{i}^{}s^{z}_i -\tilde{g} \mu_B H
\sum_{i}^{}S^z_i
\end{equation}
 where $c^{\dagger}_{i\sigma}$ is the creation operator of a
hole with spin $\sigma$ in the bound state associated with the $i$th
Mn impurity.  $\vec{R}_i$ ($i=1,N_d$) are the positions of the Mn
impurities, and  the hopping matrix $t_{ij}=t(\vert
\vec{R}_i-\vec{R}_j\vert)$ is given by $t(r) = 2\left( 1 + r/ a_B\right)
\exp{(-r/ a_B)}E_b $ \cite{Berciu2004}, where the $E_b$ is the binding
energy of the hole, and $a_B = \epsilon \hbar^2/m_h e^2$ is the
hydrogenic Bohr radius.  The second term is the antiferromagnetic exchange interaction
between Mn spins $\vec{S}(i)$ and hole spins where $\vec{s}_i=c^{\dagger}_{i\alpha} {1 \over 2} \vec{\sigma}_{\alpha\beta}
c_{i\beta}$ . The exchange integral is given by $ J_{ij}
= J \exp{ (-2 { \vert \vec{R}_i - \vec{R}_j \vert / a_B} ) }, $
which is related to the probability of finding the hole in the impurity state
around $j$ on the $i$th Mn spin.  The last two terms in Eq.  (\ref{eq1})
describe interactions of the spins of the Mn and holes with an external magnetic field H.

We perform a mean-field calculation, following the same procedure described in references [\onlinecite{Berciu2001,Berciu2004}] with carrier and Mn concentrations similar to the experimental data. In our calculations, we use $a_B=7.8\AA$, $m_h=0.5 m_e$, $E_b=112$ meV, $J=15$ meV and the lattice constant of GaAs $a=5.65\AA$. We consider three different situations of disorder: for the non-disordered case, the Mn ions form an ordered lattice. In the weak disorder case we allow the Mn ions to move to a nearest neighbor with a probability of 20$\%$. Finally, in the strong disorder limit, they are randomly distributed in the DMS. For comparison with the experiments, we use two concentrations, x=0.009 and x=0.004, where due to the compensation mechanisms, the hole density $n_h$ is a fraction of the effective Manganese density $n_{Mn}$. For all the mean-field calculations we use tridimensional lattices with 1000 Mn spins and for the disordered systems we perform avarages over 500 realizations of disorder.

\begin{figure}[b] % float placement: (h)ere, page (t)op, page (b)ottom, other (p)age
  \centering
  % file name: H:/USB MEMORY (G)/GaMnAs/paper-dms/PRB/magteo.eps
  \includegraphics[width=1.0\columnwidth,clip]{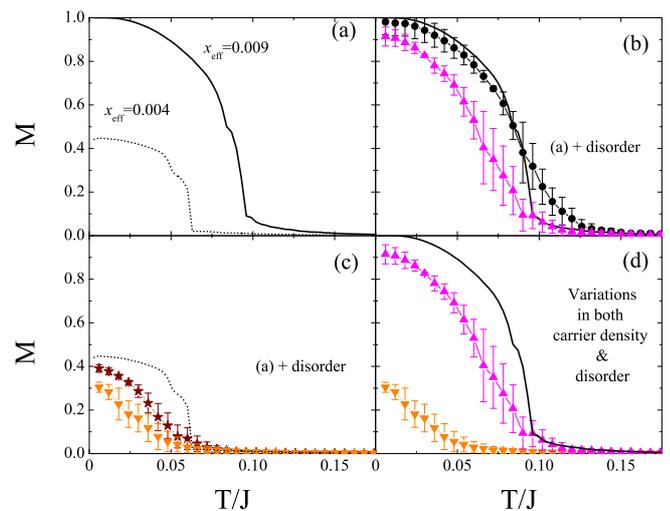}
  \caption{(a) Mean-Field Magnetization versus temperature for $x=0.009$ (solid line) and $x=0.004$ (dotted line) in the absence of disorder. (b) Magnetization versus temperature for $x=0.009$ and non-disordered (solid line), weakly disordered (up triangles) and strongly disordered (circles) systems. (c) Magnetization versus temperature  for $x=0.004$ non-disordered (dotted line), weakly disordered (stars) and strongly disordered (down triangles) systems. (d) Magnetization versus temperature for non-disordered $x=0.009$ (solid line), weakly disordered $x=0.009$ (up triangles) and strongly disordered $x=0.004$ (down triangles). All curves use $n_h/n_{Mn}=0.25$ and are normalized by the magnetization of $x=0.009$ at zero temperature. }
  \label{Fig8}
\end{figure}
We consider two different fractions, $n_h/n_{Mn}=0.2$ and $n_h/n_{Mn}=0.25$. From the experimental results, we observe that although the saturation magnetization and the shape of the magnetization curve are strongly affected by the irradiation process, the critical temperature is only slightly modified (see for example, samples A and C). 
Using this mean-field theory we can see, as illustrated in Fig.~\ref{Fig8} (a), that if we decrease the Mn concentration with a fixed ratio $n_h/n_{Mn}=0.25$, we obtain both a suppression in T$_c$ and a reduction in the saturation magnetization without a change in the shape of the magnetization curve. On the other hand, if we keep $x$ fixed and change the amount of disorder in the system, we can increase $T_c$, as is clearly seen in Fig.~\ref{Fig8} (b) for $x=0.009$. The disorder can also modify the shape of the magnetization curve, as we see in Fig.~\ref{Fig8}(c) for $x=0.004$. So if instead of just changing the Mn concentration we also increase the disorder for the lower concentration, the change in $x$ is compensated by an increase in T$_c$ due to the increase in the disorder~\cite{Berciu2001}. We can also see a change in the concavity of the curve, in agreement with the experimental data. In this sense, in Figure~\ref{Fig8}(d) we emulate the irradiation process by a suppression in the Manganese concentration and an increase in the disorder of the system. The three magnetization curves are similar to the ones obtained experimentally for non-irradiated samples, low and high ion doses and can be compared with the results shown in Fig.~\ref{Fig5}.
\begin{figure}[t]
  \includegraphics[width=0.7\columnwidth,clip]{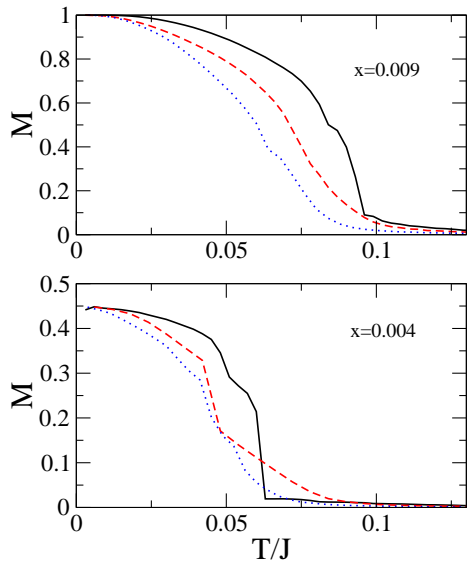}
  \caption{(a) Mean-Field Magnetization versus temperature in the absence of disorder for: (a) $x=0.009$ with $n_h/n_{Mn}=0.25$ (solid line),  $n_h/n_{Mn}=0.15$ (dashed line) and  $n_h/n_{Mn}=0.10$ (dotted line) and (b) $x=0.004$  with $n_h/n_{Mn}=0.25$ (solid line),  $n_h/n_{Mn}=0.15$ (dashed line) and  $n_h/n_{Mn}=0.10$.  All curves are normalized by the magnetization of $x=0.009$ at zero temperature.}
  \label{Fig9}
\end{figure}
If instead of increasing the disorder of the system, we consider an increase in $n_h/n_{Mn}$ after the irradiation process, we can also compensate the change in T$_c$. However, we do not obtain the correct magnetization curve shape. We could also consider that the irradiation process does not modify the Mn concentration but only decrease the carrier concentration but in this case we can understand the decrease in the critical temperature but not the strong suppression of the saturation magnetization, which does not depend on the carrier concentration at low temperatures.  In Fig.~\ref{Fig9} we support our analysis by showing the magnetization curves for both $x=0.009$ and $x=0.004$ and different carrier concentrations.

\section{Conclusions}

We irradiated GaMnAs epilayers by 100-1000 KeV light ion beams and studied the change in their transport and magnetic properties as a function of the irradiation dose. We performed in-situ sheet resistance measurements and found that irradiation by ion beams can electrically isolate a diluted magnetic semiconductor, similarly to what is seen in p and n-doped GaAs. We performed magnetic measurements on the irradiated samples and  observed a suppression of the saturation magnetization with the increase of the irradiation dose.  However, this same change was not observed in the critical temperature, which was only slightly modified. In order to better understand the roles of the change in both hole concentration and disorder of the system in the observed magnetization curves, we compared our experimental results with a mean-field calculation for the impurity band picture. 

This comparison suggests that the increase of disorder plays an important role in the magnetic properties of the DMS after irradiation.  Furthermore, our experimental results show that the increase of disorder due to an intense irradiation process leads also to the formation of magnetic clusters in the sample. We performed field-cooled and zero-field cooled magnetic measurements and found the energy barrier necessary to overcome such configuration.

We show that the irradiation of GaMnAs epilayers by 100-1000 KeV light ion beams can modify the sample in a quantitatively controlled way. This proof-of-principle experiment has interesting potential consequences for the manipulation of GaMnAs. It is important to note that, with an appropriate experimental setup, keV and MeV ion beams can be focused to few-microns-wide spots. This opens a door to, keeping the planarity of the sample, writing on the GaMnAs and creating neighboring regions with different carrier densities. It is actually possible, controlling the ion beam intensity and irradiation time during the writing process, to create regions on the sample where the carrier density varies continuously.

\section{acknowledgments}
This work was supported by the Brazilian agencies CNPq, FUJB and FAPERJ, and by L'Oreal-Brazil.

\end{document}